\documentclass[%
 reprint,
 superscriptaddress,
 amsmath,amssymb,
 aps,
]{revtex4-1}

\usepackage{graphicx}% Include figure files
\usepackage{dcolumn}% Align table columns on decimal point
\usepackage{bm}% bold math
\usepackage{xcolor}
\begin{document}

%\preprint{APS/123-QED}

\title{Exploring the Mass Radius of $^4$He and Implications for Nuclear Structure}
%% Force line breaks with \\
%%\thanks{A footnote to the article title}%

\author{Rong Wang}
\email{rwang@impcas.ac.cn}
\affiliation{Institute of Modern Physics, Chinese Academy of Sciences, Lanzhou 730000, China}
\affiliation{School of Nuclear Science and Technology, University of Chinese Academy of Sciences, Beijing 100049, China}

\author{Chengdong Han}
\email{chdhan@impcas.ac.cn}
\affiliation{Institute of Modern Physics, Chinese Academy of Sciences, Lanzhou 730000, China}
\affiliation{School of Nuclear Science and Technology, University of Chinese Academy of Sciences, Beijing 100049, China}

\author{Xurong Chen}
\email{xchen@impcas.ac.cn}
\affiliation{Institute of Modern Physics, Chinese Academy of Sciences, Lanzhou 730000, China}
\affiliation{School of Nuclear Science and Technology, University of Chinese Academy of Sciences, Beijing 100049, China}


\date{\today}% It is always \today, today,

\begin{abstract}
In this study, we determine the mass radius of $^4$He, a very light nucleus,
by examining the near-threshold $\phi$-meson photoproduction data of the LEPS Collaboration.
To assess the gravitational form factor of $^4$He, we employ multiple models
for the mass distribution, including Yukawa-type, exponential, Gaussian, and uniform functions.
The mass radius of $^4$He is measured to be $1.70\pm0.14$ fm,
which is approximately equal to the charge radius of $^4$He.
Surprisingly, in contrast to the findings of the proton,
no noticeable discrepancy between the charge radius
and the mass radius is noted for the $^4$He nucleus.
The proton and neutron distributions within $^4$He are likely to be identical,
confirming its regular tetrahedral structure in a new way.
We propose exploring the difference between charge and mass radii
as a new approach to examine the nuclear structure.
\end{abstract}

%% \keywords{Suggested keywords}%Use showkeys class option if keyword
\maketitle

%\tableofcontents

\section{Introduction}
\label{sec:intro}

The mass radius is an important and basic property for any composite system,
from the sub-atomic particles of very small scale in high energy physics
to the galaxies of very large scale in astrophysics.
The trace anomaly from the quantum corrections of quantum chromodynamics (QCD)
breaks the conformal symmetry \cite{Nielsen:1977sy,Adler:1976zt,Collins:1976yq},
and it is one key mechanism for the nucleon mass generation
\cite{Ji:1994av,Ji:1995sv,Ji:2021pys,Ji:2021mtz,Lorce:2021xku,Wang:2019mza,Kou:2021bez}
and responsible for the most of the mass of the visible universe.
The mass of a particle can be viewed as the response of the particle
to an external gravitational field.
The gravitational form factors (GFF) of a particle are defined
as the off-forward matrix elements
of energy-momentum tensor (EMT) in the particle state \cite{Pagels:1966zza,Teryaev:2016edw,Polyakov:2018zvc}.
The GFFs contain the fundamental properties of the particle,
such as the mass and spin \cite{Teryaev:2016edw,Polyakov:2018zvc}.
In the Breit frame, the 00-component of the static EMT is the energy density,
and the energy density of the whole system should be normalized to the mass \cite{Polyakov:2018zvc}.
Therefore the mass density distribution and mass radius are all defined
and derived from the GFFs.

The naive way to probe GFFs is via graviton scattering,
however it is infeasible due to the weakness of gravitational
interaction of a particle.
A practical opportunity is via the measurement of generalized
parton distributions (GPD) from various exclusive scattering processes.
The second Mellin moments of GPDs yield the combinations of GFFs \cite{Teryaev:2016edw,Polyakov:2018zvc}.
Recently, with some QCD analyses, it is suggested that the diffractive vector-meson
photoproduction near the production threshold is sensitive to
the gluonic GFFs of the target
\cite{Kharzeev:2021qkd,Wang:2021dis,Guo:2021ibg,Frankfurt:2002ka,Mamo:2019mka,
Hatta:2018ina,Hatta:2018sqd,Hatta:2019lxo,Mamo:2022eui,Ji:2020bby,Sun:2021gmi}.
These QCD analyses result in three approaches for calculating the scattering amplitude
of near-threshold vector-meson photoproduction: GPD approach \cite{Guo:2021ibg,Frankfurt:2002ka},
holographic QCD approach \cite{Mamo:2019mka,Hatta:2018ina,Hatta:2018sqd,Hatta:2019lxo,Mamo:2022eui},
and the factorization based on the vector-meson-dominance (VMD) model \cite{Kharzeev:2021qkd,Wang:2021dis}.

In experiment, the determination of mass radius is closely
related to the extractions of GFFs from experimental data.
Actually there are some pioneering works in determining
the mass radii and GFFs of the proton \cite{Kharzeev:2021qkd,Wang:2021dis,Duran:2022xag},
the deuteron \cite{Wang:2021ujy} and the pion \cite{Kumano:2017lhr,Xu:2023bwv}.
From these analyses, the mass radii of the studied hadronic
particles are all smaller than the electric charge radii.
With the recent experimental data of near-threshold J/$\Psi$
photoproduction at Jefferson Lab (JLab), physicists have extracted
the gluonic gravitational form factors of the proton with
both the GPD approach and the holographic approach.
They found that the mass radius is notably smaller than
the charge radius, and the proton structure consists
of three distinct regions \cite{Duran:2022xag}.

In principle the concepts of mass radius and GFFs can be
applied to a large hadronic system.
At low energy, the nucleonic degree of freedom is valid
for describing the static properties
and low-energy reactions of an atomic nucleus.
However, at high energy and a more fundamental level,
the nucleus is built with quarks and gluons.
It is very interesting to find out whether there is
the difference between the mass radius and the charge radius
of a nucleus. From our previous analysis,
the mass radius of the loosely bound deuteron is slightly smaller
than its charge radius \cite{Wang:2021ujy}. However for the tightly bound nucleus,
such as the $^4$He, we still lack the information
on its mass radius and the related analysis.

The charge radius of $^4$He is precisely measured
to be $1.67824(83)$ fm with the technique of muon-atom spectroscopy \cite{Krauth:2021foz},
and the world average from electron elastic scattering experiments
is $1.681(4)$ fm \cite{Sick:2008zza}. An older combined analysis gave
the average charge radius of $^4$He to be $1.6755(28)$ fm \cite{Angeli:2013epw}.
However the mass radius of $^4$He has never been studied.
In this work, we investigate the mass radius of the helium nucleus
from an analysis of the $|t|$-dependence of the differential cross section
of near-threshold $\phi$-meson photoproduction,
which could provide important information about
the gravitational properties and the internal structure
of a large hadronic system, especially the transverse spatial distributions.

\section{Near-threshold $\phi$-meson photoproduction and gravitational form factors}
\label{sec:phi-production-and-GFFs}

The GFFs are the matrix elements of the EMT,
which encode the mechanical properties of a composite particle.
The trace anomaly of EMT sets up a mass scale of the hadronic system,
and it is one key component of the origin of the proton mass according to the QCD analysis
\cite{Ji:1994av,Ji:1995sv,Ji:2021pys,Ji:2021mtz,Lorce:2021xku,Wang:2019mza,Kou:2021bez}.
In the chiral limit, the scale anomaly is shown in the trace of EMT
of QCD in terms of scalar gluon operator \cite{Kharzeev:2021qkd}.
The trace anomaly in QCD is the pure quantum effect from gluon fluctuations.
In Kharzeev's view, in the weak gravitational field,
the trace of EMT and the temporal component of EMT $T^{00}$ coincide \cite{Kharzeev:2021qkd}.
The scalar GFF is then defined by Kharzeev as the form factor of the trace of the EMT.
It is lorentz-invariant and defines the mass distribution of the system.
In the chiral limit of massless quarks, the information about the mass radius
of a hadronic system is contained in the matrix element of scalar gluon
operator at a nonzero momentum transfer, for the matrix element does not
depend on the strong coupling constant due to the scale anomaly \cite{Kharzeev:2021qkd}.

In the nonrelativistic limit and based on the VMD model,
the amplitude of a vector meson photoproduction can be safely
factorized into a short-distance part describing the electric polarizability
of the $q\bar{q}$ pair, and the matrix element of
the chromoelectric operator over a hadron \cite{Kharzeev:2021qkd}.
The scalar part in the chromoelectric operator is the trace of the EMT,
and it dominates near the threshold of the vector-meson
photoproduction as a consequence of scale anomaly.
Therefore the vector meson photoproduction amplitude can be expressed as \cite{Kharzeev:2021qkd},
\begin{equation}
\begin{split}
    \mathcal{M}_{\gamma p \rightarrow \phi p^{\prime}}(t)
    = -Qec_2\frac{16\pi^2M}{b} \left<p^{\prime}|T_{\mu}^{\mu}|p\right>.
\end{split}
\label{eq:scattering-amplitude}
\end{equation}
The differential cross section is then computed with the square
of the scattering amplitude, which is written as,
\begin{equation}
\begin{split}
    \frac{d\sigma_{\gamma p \rightarrow \phi p^{\prime}}}{dt}
    = \frac{1}{64\pi s}\frac{1}{|E_{\gamma,~c.m.}|^2}
    \big| \mathcal{M}_{\gamma p \rightarrow \phi p^{\prime}}(t) \big|^2.
\end{split}
\label{eq:diff-xsection}
\end{equation}
With the above analysis and Eq. (\ref{eq:scattering-amplitude}),
the differential cross section is proportional to the square
of the scalar GFF of the hadronic target, as,
\begin{equation}
\begin{split}
    \frac{d\sigma_{\gamma p \rightarrow \phi p^{\prime}}}{dt} \propto |G(t)|^2.
\end{split}
\label{eq:diff-xsection-simple}
\end{equation}
In some sense and the phenomenological view, the theoretical structures of
the GFFs are similar in the processes probed by the graviton and the $V^{*}V$ in the VMD model.

In this work, we study the $|t|$-dependence of the differential
cross section of $\phi$-meson photoproduction off the $^4$He target
with the theoretical framework in terms of the scalar GFF discussed above,
in order to extract the mass radius of $^4$He.
For the convenience of discussions, we may define a normalized scalar GFF $F(t)$ as,
\begin{equation}
\begin{split}
    F(t) = \frac{G(t)}{M}.
\end{split}
\label{eq:reduced-scalar-GFF}
\end{equation}
The mass radius then can be simply computed with the slope
of the scalar GFF at zero momentum transfer ($t=0$ GeV$^2$), as,
\begin{equation}
\begin{split}
    \left<r_{\rm m}^2\right> = - 6\frac{dF(t)}{dt} = -\frac{6}{M}\frac{dG(t)}{dt},
\end{split}
\label{eq:mass-radius}
\end{equation}
which is also discussed in the following section.
To be consistent with our previous analyses of the mass radii
of the proton and the deuteron \cite{Wang:2021dis,Wang:2021ujy},
we apply the same theoretical
framework of the scalar GFF discussed above.

\section{Various density distributions and form factors}
\label{sec:form-factor-models}

The root-of-mean-square (RMS) radius $\sqrt{\left<r^2\right>}$
from a density distribution $\rho(r)$ is defined as,
\begin{equation}
\begin{split}
    \left<r^2\right> = \int_0^{\infty} r^2 \rho(r) 4\pi r^2 dr.
\end{split}
\label{eq:rms-radius-def}
\end{equation}
In the low-momentum elastic scattering process, the form factor $F(q)$
of the target is measured, and it is the Fourier transformation
of the density distribution $\rho(r)$.
For a continuous density distribution and under the small momentum exchange,
the RMS radius also can be easily computed with the slope
of the form factor at $Q^2=0$ GeV$^2$, which is written as,
\begin{equation}
\begin{split}
    \left<r^2\right> = -6\frac{dF(q^2)}{dq^2}\bigg|_{q^2=0}.
\end{split}
\label{eq:rms-radius-calculation}
\end{equation}

For different hadronic systems, the density distributions are different.
The various and typical density distributions,
the corresponding form factors and RMS radii are listed in Table \ref{tab:density-FF-radius}.
For the light meson, such as the pion, the density reduces quickly with
the increasing radial distance, and the density distribution is taken as the Yukawa potential form.
The corresponding form factor of the pion is monopole-like.
The dipole form factor from exponential distribution describes
well the form factor of the proton in a wide kinematical range.
For the heavy nucleus, such as the lead nucleus, the density distribution is approximately
described with the uniform distribution or the Fermi distribution
due to the saturation property of nuclear matter.
The $^4$He is a light and compact nucleus.
The density distribution and the form factor of $^4$He
should be different from those of the proton and the heavy nucleus.

\begin{table}[h]
    \caption{Some density distributions, the corresponding form factors and RMS radii.}
        \renewcommand\arraystretch{2}
		\begin{tabular}{cccc}
        \hline\hline
            Model        &           $\rho(r)$                      &       $F(q)$                               & $\sqrt{\left<r^2\right>}$     \\
        \hline
          Point-like   & $\frac{1}{4\pi r^2}\delta(r)$            & 1                                          & 0                              \\
          Yukawa-type  & $\frac{\Lambda^2}{4\pi r}e^{-\Lambda r}$ & $\frac{1}{1+q^2/\Lambda^2}$                & $\sqrt{\frac{6}{\Lambda^2}}$   \\
          Exponential  & $\frac{\Lambda^3}{8\pi }e^{-\Lambda r}$  & $\frac{1}{\left(1+q^2/\Lambda^2\right)^2}$ & $\sqrt{\frac{12}{\Lambda^2}}$  \\
          Gaussian     & $\left(\frac{\Lambda^2}{\pi}\right)^{3/2}e^{-\Lambda^2r^2}$ & $e^{-q^2/(4\Lambda^2)}$ & $\sqrt{\frac{3}{2\Lambda^2}}$  \\
          Uniform      & $\frac{3}{4\pi R^3}\theta(R-r)$          & $\frac{3j_1(qR)}{qR}$                      & $\sqrt{\frac{3R^2}{5}}$        \\
        \hline\hline
		\end{tabular}
    \label{tab:density-FF-radius}
\end{table}

In this work, our goal is to determine the RMS mass radius
of the $^4$He nucleus from the coherent and diffractive scattering off the target.
Thus the density distribution discussed above is specifically
the mass distribution, and the form factor is the scalar GFF.
To see which model of mass distribution and scalar GFF describes well the $^4$He nucleus,
the experimental data of the near-threshold $\phi$-meson photoproduction off $^4$He 
are fitted with various function forms of the scalar GFF.

\section{Data analysis and results}
\label{sec:analysis-and-results}

\begin{figure*}[htbp]
\begin{center}
\includegraphics[width=0.85\textwidth]{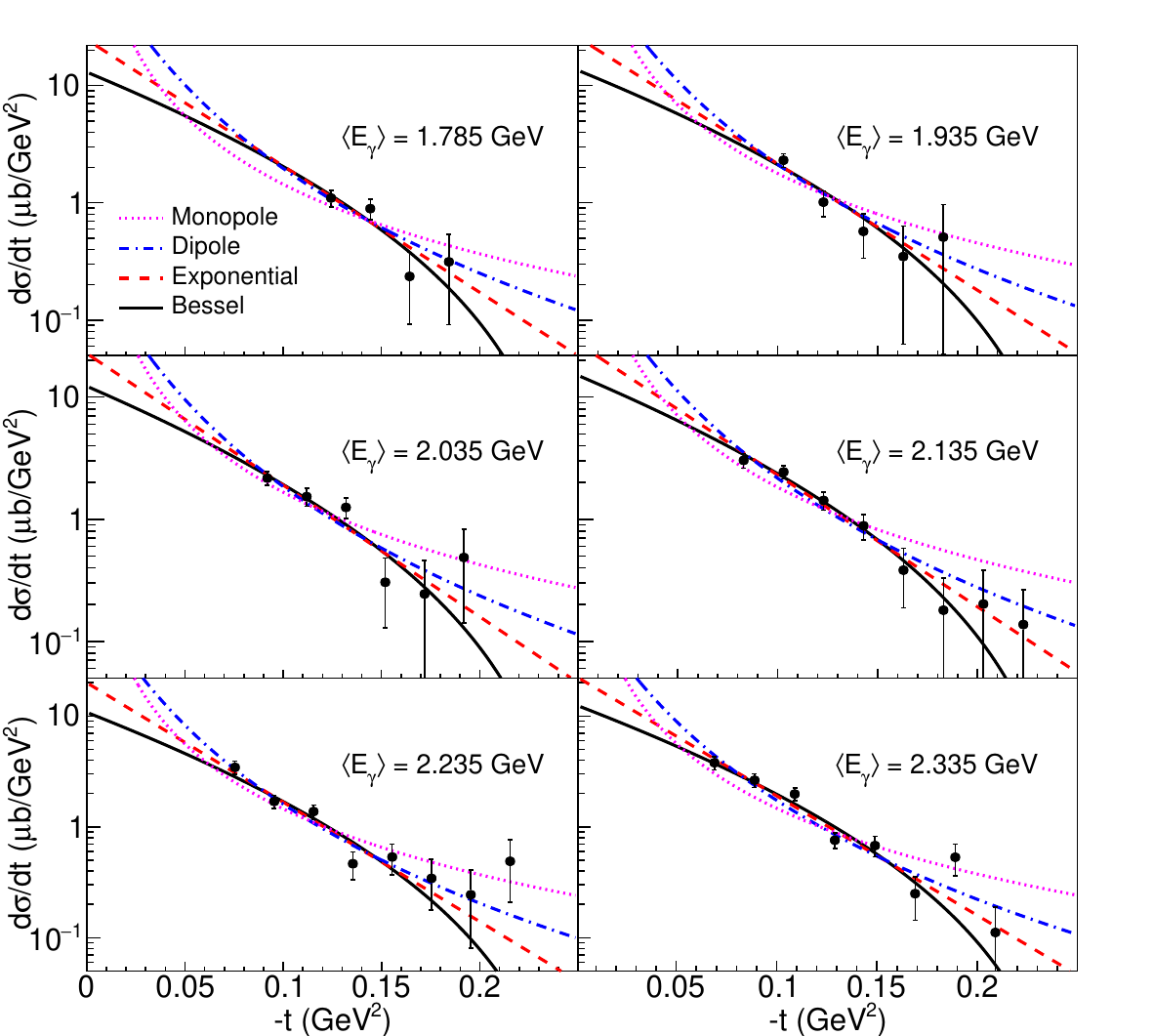}
\caption{
  (color online) The measured differential cross sections of the coherent
  $\phi$-meson photoproduction off the $^{4}$He nucleus near the threshold,
  compared with various models for the nuclear GFF.
  The experimental data are taken from LEPS Collaboration's publication \cite{LEPS:2017nqz}.
  The magenta dotted curves show the fitting results of the monopole GFF from Yukawa-type mass distribution.
  The blue dash-dotted curves show the fitting results of the dipole GFF from exponential mass distribution.
  The red dashed curves show the fitting results of the exponential GFF from Gaussian mass distribution.
  The black solid curves show the fitting results of the Bessel GFF from uniform mass distribution.
}
\label{fig:cross-sections-and-models}
\end{center}
\end{figure*}

Fig. \ref{fig:cross-sections-and-models} shows the measured differential cross
sections of the near-threshold $\phi$ meson photoproductions as a function of $t$
at different energies from LEPS collaboration \cite{LEPS:2017nqz}.
For the momentum transfer $\tilde{t}$ in the LEPS data,
$|t|_{\rm min}$ is subtracted. In this analysis, we remove the correction
on the momentum transfer by calculating the $|t|_{\rm min}$'s
of the reaction $\gamma ~{\rm ^4He} \rightarrow \phi ~{\rm ^4He}$
at different photon energies.
The differential cross sections are fitted with the models
of various function forms for the scalar GFF.
We investigated four different models: the monopole model, the dipole model,
the Gaussian mass distribution model, and the uniform mass distribution model.
In the data fitting with each model, the scalar GFF is the same for all the experimental data
at different photon energies, and only the normalizations are different at different energies.
One sees that all the models reproduce
the experimental data in the narrow $|t|$ range.
Nevertheless, the scalar GFF from Gaussian mass distribution
most agree with the differential cross sections.
More and precise experimental data in a large region of kinematic
$|t|$ are needed to differentiate the models more clearly.

To quantify the quality of fit, the reduced $\chi^2$ are calculated
for different models, which are listed in Table \ref{tab:fitting-results}.
One finds that the Gaussian mass distribution model describes the experimental data the best
with the smallest $\chi^2/N_{\rm dof}$ quite close to 1.0.
For describing a compact and small nucleus,
the Gaussian distribution of the mass is a rather good choice.
The uniform distribution of the mass also fits well
the experimental data with $\chi^2/N_{\rm dof}<1.2$.
But the uniform mass distribution is just an over-ideal distribution
for a heavy nucleus with the perfect nuclear saturation property.

The final results of the model fittings are summarized in Table \ref{tab:fitting-results},
including the extracted slope parameter $\Lambda$ or $R$ in the modeled GFFs,
and the related mass radii.
In the least-square fit of each model, the slope parameter $\Lambda$ (or $R$)
of the GFF is the same for all the cross-section data at different photon energies.
It is clearly shown that the extracted mass radii under different model assumptions vary significantly.
There is a strong model-dependence of the extracted mass radius.
The first reason is that the $|t|$ range covered
by the experimental data is narrow, about 0.15 GeV$^2$.
The second reason is that the effective extrapolation of the slope to $t=0$ GeV$^2$ requires more
experimental data at small $|t|$ close to zero.
On the other side, the model-dependence is very natural in extraction of mass radius,
since any model assumption definitely introduces the model uncertainty.
Nonetheless, based on the current limited data, the exponential GFF
of Gaussian mass distribution is the most effective model tested.

In the model fittings, the normalizations at different energies are set as free parameters,
for we can not precisely or accurately compute them under the fundamental theory so far.
Thus, including the parameter for modeling the scalar GFF, there are seven free parameters.
As all the free parameters are important and have definite physical meanings,
the multi-parameter confidence region is considered in the analysis.
Therefore we apply $\Delta\chi^2=8.38$ for the error estimations
in multi-dimensional parameter space at the confidence level of 70\%,
suggested by the `minuit' manual.

\begin{table}[h]
    \caption{The determined model parameters, the extracted mass radii
             and the fitting qualities $\chi^2/N_{\rm dof}$ with various models
             for the mass distribution.}
        \renewcommand\arraystretch{1.5}
		\begin{tabular}{cccc}
        \hline\hline
            Model        & $\Lambda$ (GeV) &   $\sqrt{\left<r^2\right>}$ (fm)   &  $\chi^2/N_{\rm dof}$   \\
        \hline
          Yukawa-type  & $0.045 \pm 0.051$ &   $10.72 \pm 12$   &   $93.80/32$     \\
          Exponential  & $0.220 \pm 0.063$ &   $3.10 \pm 0.89$    &   $40.98/32$      \\
          Gaussian     & $0.142 \pm 0.011$ &   $1.70 \pm 0.14$    &   $34.09/32$      \\
                         & $R$ (GeV$^{-1}$)  &           &          \\
          Uniform      & $8.97 \pm 0.47$   &   $1.37 \pm 0.08$    &   $37.50/32$      \\
        \hline\hline
		\end{tabular}
    \label{tab:fitting-results}
\end{table}

\section{Discussions and summary}
\label{sec:discussions-and-summary}

\begin{figure}[htbp]
\begin{center}
\includegraphics[width=0.46\textwidth]{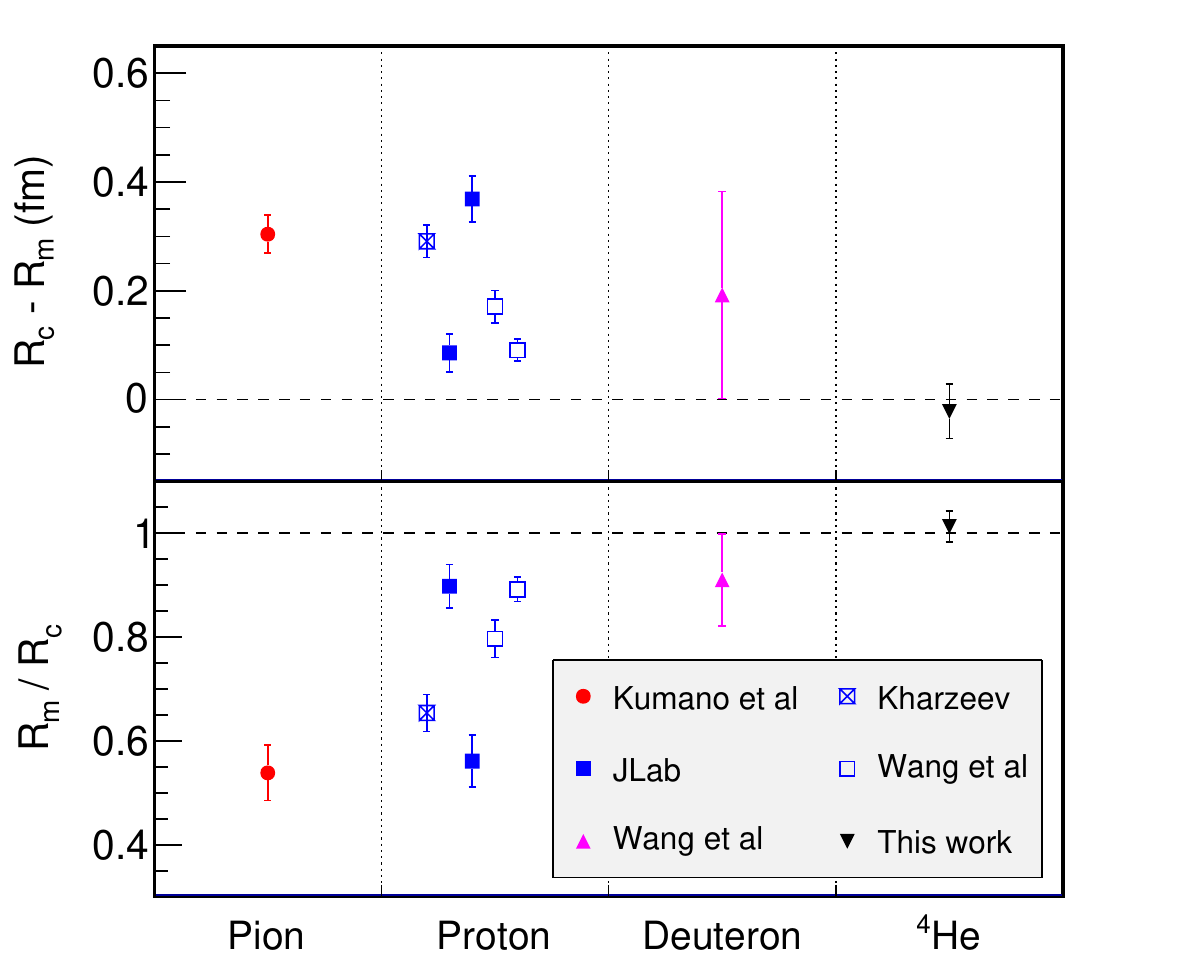}
\caption{
  (color online) The determined mass radii of the pion, the proton,
  the deuteron, and the $^4$He nucleus, from various groups.
  The pion result is taken from Ref \cite{Kumano:2017lhr}.
  The proton results are taken from Refs. \cite{Kharzeev:2021qkd,Wang:2021dis,Duran:2022xag}.
  The two values by JLab Hall C Collaboration are from the holographic
  QCD approach and the GPD approach.
  The two values by Wang et. al. are from the analysis of
  only $\phi$-photoproduction data and the combined analysis of the
  photoproductions of $\omega$, $\phi$, and J/$\psi$ mesons.
  The deuteron result is taken from Ref. \cite{Wang:2021ujy}.
  The helium result is from this work with the Gaussian mass distribution model.
}
\label{fig:mass_radii_comparisons}
\end{center}
\end{figure}

\begin{figure}[htbp]
\begin{center}
\includegraphics[width=0.42\textwidth]{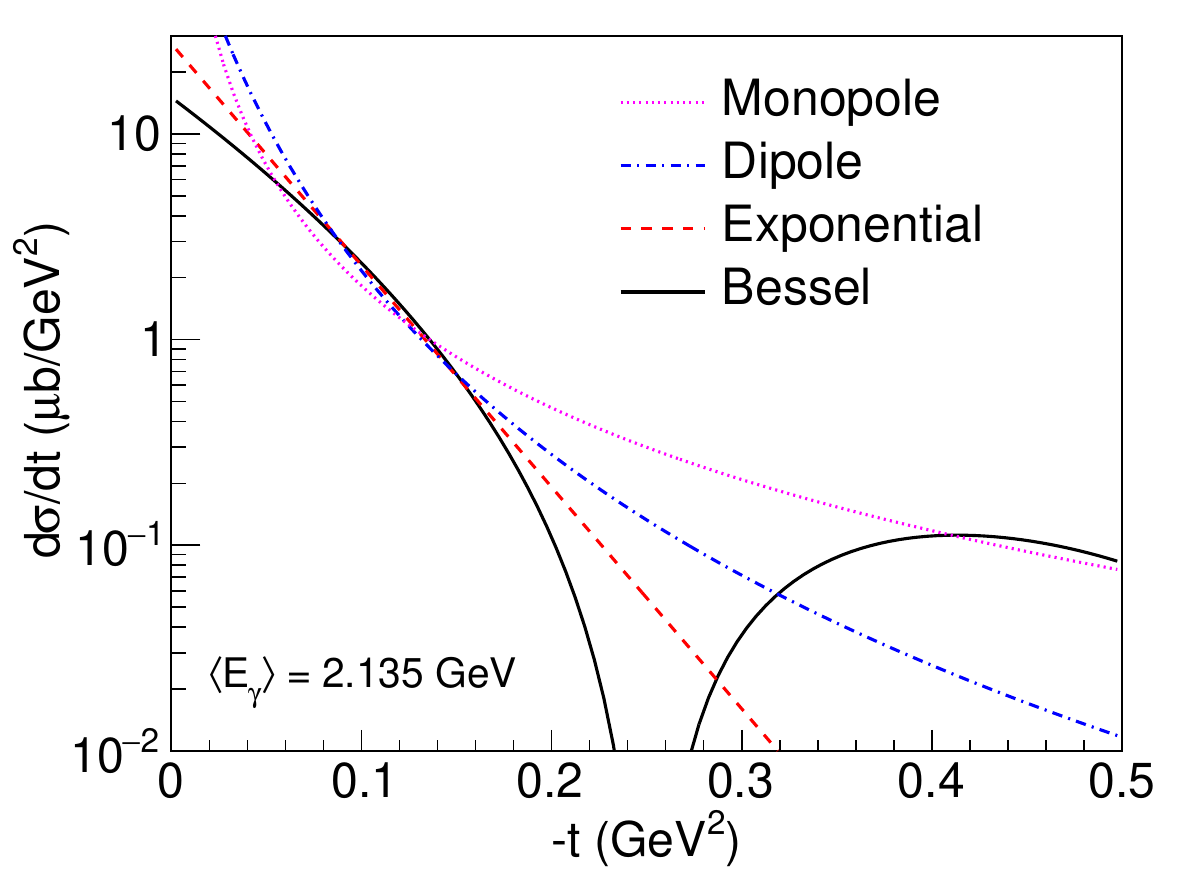}
\caption{
  (color online) The predictions of the differential cross section
  of $\phi$ meson photoproduction near threshold in a broad $t$ range
  with different models for the scalar GFF.
  The magenta dotted curve shows the prediction of monopole GFF.
  The blue dash-dotted curve shows the prediction of dipole GFF.
  The red dashed curve shows the prediction of exponential GFF.
  The black solid curve shows the prediction of Bessel-type GFF.
}
\label{fig:diff_xsection_predictions}
\end{center}
\end{figure}

From the analysis, we find that the differential cross section of
near-threshold $\phi$-meson photoproduction off the $^4$He target can be
described well at small $|t|$ with the exponential GFF of Gaussian mass distribution.
The $^4$He is a light nucleus, thus the uniform distribution is
an inappropriate approximation for its mass distribution.
Meanwhile the $^4$He is a compact nucleus,
its density in the center should not change fast
with the radial distance increasing.
This is probably why the Gaussian distribution model is most
consistent with the measured differential cross sections.

For a heavy nucleus, the density is fairly a constant in the central
region due the saturation property of nuclear force,
and the density distribution can be approximately modeled
with a uniform distribution, or the Fermi model
with a radius parameter and a surface thickness \cite{Hofstadter:1956qs,Hofstadter:1957wk}.
For $^{12}$C, the best fit of the electron elastic scattering data
was found lying between a Gaussian model and a uniform model \cite{Hofstadter:1956qs}.
The latter studies found that the charge distributions of
$^{12}$C and $^{16}$O can be described well with the harmonic-shell charge distributions \cite{Hofstadter:1957wk}.
For the charge distribution of the very light $^4$He,
it was found that the Gaussian model is the best in the
low momentum transfer region ($<6.2$ fm$^{-2}$) \cite{Hofstadter:1956qs,Hofstadter:1957wk,Frosch:1967pz}.
The momentum transfer $|t|$ is below 6.2 fm$^{-2}$ for the LEPS data used in this analysis.
The pure Gaussian shape of charge distribution of
$^4$He is also provided by the shell model \cite{Hofstadter:1957wk}.
In this work, we find that the Gaussian distribution is also
the best model in explaining the mass distribution of $^4$He.
The shapes of the mass and charge distributions are just alike for the nucleus.

The mass radius is obtained to be $1.70\pm 0.14$ fm
based on the Gaussian mass distribution model.
The mass radius of
the helium nucleus is nearly as same as its charge radius.
This conclusion is quite surprising,
as it violates what have been found for the proton
\cite{Kharzeev:2021qkd,Wang:2021dis,Duran:2022xag},
the deuteron \cite{Wang:2021ujy} and the pion \cite{Kumano:2017lhr,Xu:2023bwv}.
Why the mass radius of the proton is much smaller than
its charge radius is a complicated and unanswered question.
Why the mass radius of $^4$He is almost the same of
its charge radius is another astounding puzzle,
which should be further investigated in the future experiments.
One may simply assume the underlying confinement mechanisms for
the proton and the nuclei are different,
which should be carefully studied with the nonperturbative QCD theory in the future.

Fig. \ref{fig:mass_radii_comparisons} shows the differences and ratios
between mass radius and charge radius, for some hadrons and nuclei examined recently.
Based on the current analyses, the differences between mass and charge radii are similar
for the pion and the proton, approximate 0.2 fm.
Nonetheless, the difference in the mass and charge radii appears to vanish in $^4$He.
As the target size increases, the ratio of mass radius to charge radius goes up approaching one.
In a picture of nucleonic degrees of freedom, the compatibility of the mass and charge radii
of $^4$He supports the regular tetrahedron-like structure of $^4$He.
Nuclear mass radius is related to both the proton and the neutron distributions inside a nucleus,
whereas nuclear charge radius primarily connect to the proton distribution.
Therefore, the radii of the proton distribution and the neutron
distribution in $^4$He are closely similar.

We argue that it is a pioneering and pivotal approach to scrutinize the nuclear structure
via the examination of the difference between the charge and mass radii of a nucleus.
For $^9$Be of the dumbbell-like structure with a neutron at the center,
its charge radius should obviously surpass its mass radius.
For $^{208}$Pb with the neutron skin on the crust,
its charge radius is anticipated to be smaller than its mass radius.
Extensive and systematical investigations on the mass and charge radii of nuclei
from different probes are highly promoted, which are essentially beneficial
for solving the puzzles on the mass radius of $^4$He
and the complex structures of nuclei.

Lastly, our analysis reveals that different models produce very different
extrapolation results on the mass radius, as shown by the disparate slopes
approaching $t=0$ GeV$^2$ in Fig. \ref{fig:cross-sections-and-models}.
To differentiate between the various forms of GFF for the light nucleus,
we recommend experimental measurements across a wide kinematic range of $|t|$.
In Fig. \ref{fig:diff_xsection_predictions}, we show the predicted differential
cross sections for $\phi$ meson photoproduction over a broad range
of $|t|$ up to 0.5 GeV$^2$, based on the fitted scalar GFFs discussed earlier.
The figures indicate that the shapes of the differential cross sections
in different models vary significantly in the $|t|$ range
above 0.2 GeV$^2$ or below 0.05 GeV$^2$.
As the cross section decreases rapidly with increasing $|t|$,
we will require high-luminosity experiments to collect coherent
and diffractive data in the $|t|$ range from 0.2 GeV$^2$ to 0.5 GeV$^2$.

The US Electron-Ion Collider (EIC) under the ongoing construction \cite{AbdulKhalek:2021gbh,Accardi:2012qut}
and the proposed Chinese Electron-Ion Collider (EicC) \cite{Anderle:2021wcy,Chen:2020ijn}
show promises in achieving this goal by utilizing the plentiful quasi-real photon flux.
These facilities will offer an ample kinematical coverage and high statistics,
and the high center-of-mass energies of the collisions will enable measurements
of near-threshold heavy quarkonium (J/$\Psi$ and $\Upsilon$) photoproductions.
Therefore, we recommend the comprehensive studies of near-threshold
vector meson photoproductions ($\phi$, J/$\Psi$ and $\Upsilon$) off nuclear targets at EIC and EicC.
Such studies will differentiate between different scalar GFF models
and reveal the puzzling mass radii and structures of the nuclei.

\begin{acknowledgments}
We thank Dr. Qin-Tao SONG for the fruitful discussions and comments.
This work is supported by the National Natural Science Foundation of China
under the Grant NOs. 12005266 and 12305127,
the International Partnership Program of the Chinese Academy of Sciences under the Grant NO. 016GJHZ2022054FN,
and the Strategic Priority Research Program of Chinese Academy of Sciences under the Grant NO. XDB34030301.
\end{acknowledgments}

\bibliographystyle{apsrev4-1}
\bibliography{refs}

\end{document}